\documentclass[aps,noshowpacs,twocolumn,superscriptaddress]{revtex4-1}  
\usepackage{graphicx}  
\usepackage{dcolumn}   
\usepackage{bm}        
\usepackage{ucs}
\usepackage{amssymb}   
\usepackage{amsmath}
\usepackage{amsfonts}
\usepackage{gensymb}
\usepackage{bibunits}
\usepackage{siunitx}
\usepackage{braket} 
\usepackage{caption,setspace}

\defaultbibliography{flatten_npl_sc}
\defaultbibliographystyle{ieeetr}

\begin{document}



\title{Strong exciton-photon coupling with colloidal nanoplatelets in an open microcavity}
%

%
\author{L.C.~Flatten} \email{lucas.flatten@materials.ox.ac.uk} \affiliation{Department of Materials, University of Oxford, Oxford OX1 3PH, United Kingdom}
\author{S.~Christodoulou} \affiliation{Nanochemistry Department, Istituto Italiano di Tecnologia, Via Morego 30, IT-16163 Genova, Italy} \affiliation{Department of Physics, University of Genoa, Via Dodecaneso 33, IT-16146 Genova, Italy}
\author{R.K.~Patel} \affiliation{Department of Materials, University of Oxford, Oxford OX1 3PH, United Kingdom}
\author{A.~Buccheri} \affiliation{Department of Materials, University of Oxford, Oxford OX1 3PH, United Kingdom}
\author{D.M.~Coles} \affiliation{Department of Materials, University of Oxford, Oxford OX1 3PH, United Kingdom} \affiliation{Atomic \& Laser Physics, Clarendon Laboratory, University of Oxford, OX1 3PU, UK}
\author{B.P.L.~Reid} \affiliation{Atomic \& Laser Physics, Clarendon Laboratory, University of Oxford, OX1 3PU, UK} 
\author{R.A.~Taylor} \affiliation{Atomic \& Laser Physics, Clarendon Laboratory, University of Oxford, OX1 3PU, UK} 
\author{I.~Moreels} \affiliation{Nanochemistry Department, Istituto Italiano di Tecnologia, Via Morego 30, IT-16163 Genova, Italy} 
\author{J.M.~Smith} \affiliation{Department of Materials, University of Oxford, Oxford OX1 3PH, United Kingdom}

\vskip 0.25cm
\date{\today}

\begin{bibunit}

\begin{abstract}
Colloidal semiconductor nanoplatelets exhibit quantum size effects due to their thickness of only  few monolayers, together with strong optical band-edge transitions facilitated by large lateral extensions. In this article we demonstrate room temperature strong coupling of the light and heavy hole exciton transitions of CdSe nanoplatelets with the photonic modes of an open planar microcavity. Vacuum Rabi splittings of $66 \pm \SI{1}{\milli\electronvolt}$ and $58 \pm \SI{1}{\milli\electronvolt}$ are observed for the heavy and light hole excitons respectively, together with a polariton-mediated hybridisation of both transitions. By measuring the concentration of platelets in the film we compute the transition dipole moment of a nanoplatelet exciton to be $\mu = (575\pm110)$~D. The large oscillator strength and fluorescence quantum yield of semiconductor nanoplatelets provide a perspective towards novel photonic devices, combining polaritonic and spinoptronic effects.
\end{abstract}

\maketitle
Strong light-matter coupling results in the formation of new eigenstates, called polaritons, that consist of an admixture of the uncoupled states. Such coupling, well known from atomic physics, was first demonstrated for an excitonic solid state system at cryogenic temperatures in GaAlAs/AlAs quantum wells \cite{weisbuch_observation_1992}. More recently room temperature polariton formation in organic molecules \cite{lidzey_strong_1998,agranovich_cavity_2003} and two-dimensional materials \cite{dufferwiel_exciton-polaritons_2015,liu_strong_2015} has been achieved. Strong coupling of colloidal nanocrystals in the form of quantum dots has been previously demonstrated in evanescently coupled bilayer microcavities \cite{giebink_strong_2011}. The incorporation of polaritons in semiconducting devices allows intriguing technological advances such as threshold-less polariton lasing \cite{christopoulos_room-temperature_2007, kena-cohen_room-temperature_2010}, the formation of polariton solitons \cite{sich_observation_2012} and a feasible route to interesting phenomena such as polariton condensation, superfluidity and vortex formation \cite{amo_superfluidity_2009, byrnes_exciton-polariton_2014}. \\
Most of these systems require cryogenic temperatures and complex device fabrication. Recent advances in nanoparticle synthesis have reinforced the prospect for improving on these systems, allowing more facile polariton implementations even at room temperature. Prominent examples are colloidal nanoparticles, which are the subject of active research due to their relatively simple chemical preparation and precisely tunable spectral properties \cite{ekimov_quantum_1985,yin_colloidal_2005}. Zero-dimensional quantum dots \cite{guyot-sionnest_colloidal_2008} and one-dimensional nanorods\cite{carbone_synthesis_2007} are ubiquitous in the field of nanophotonics. Interestingly, progress in chemical synthesis have allowed the preparation of colloidal, atomically-thin, quasi-two-dimensional nanoplatelets (NPLs) \cite{ithurria_quasi_2008} with remarkable spectral properties such as narrow absorption and emission linewidths at room temperature ($<\SI{40}{\milli\electronvolt}$) and quantum yields close to 50$\%$ \cite{ithurria_colloidal_2011,tessier_spectroscopy_2012}. The large coherence area of the exciton causes an increased oscillator strength and fast exciton recombination rates down to picosecond values, exceeding that observed in traditional quantum dots by three orders of magnitude \cite{naeem_giant_2015}. Indeed, it is this large oscillator strength and the large exciton binding energy of $R_{\rm{exc}} \approx 100 - \SI{300}{\milli\electronvolt}$ \cite{naeem_giant_2015} that makes semiconducting nanoplatelets good candidates for entering the strong exciton-photon coupling regime at room temperature. \\
\begin{figure*}
\includegraphics[width=0.9\textwidth]{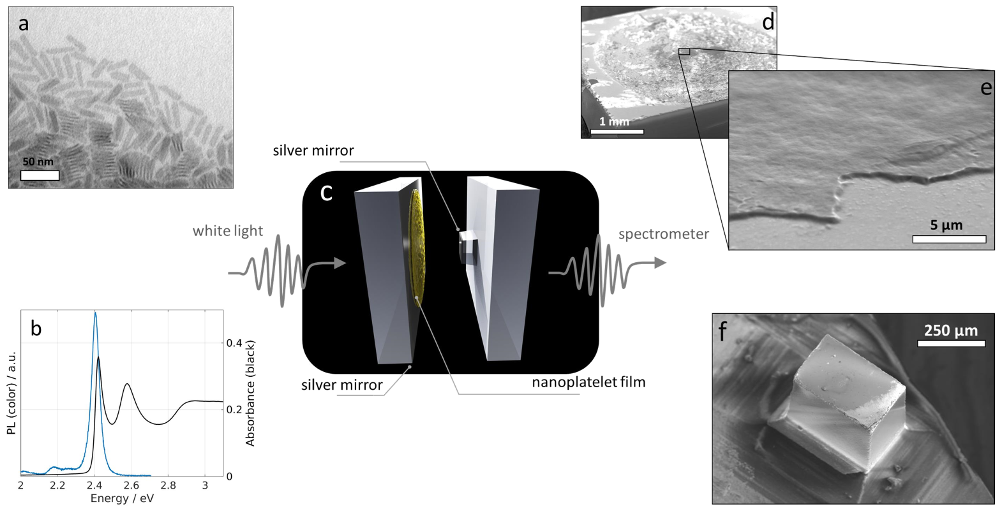}
\caption{CdSe nanoplatelets: a) TEM image of dispersed CdSe nanoplatlets. Some stacked platelets are visible towards the bottom laying sideways. b) Photoluminescence (color) and absorbance (black) of a CdSe nanoplatelet film deposited by dropcasting from solution. c) Experimental setup: Open-access microcavity consisting of two opposing, freely positionable silver mirrors. d) SEM micrograph of silver mirror (left side of cavity) with e) dropcast nanoplatelet film. f) SEM micrograph of fused silica plinth coated with $\SI{50}{\nano\meter}$ of silver (right side of cavity).\label{fig1} }
\end{figure*}
Here we give the first experimental evidence for polariton formation at room temperature with quasi two-dimensional NPLs. We use a planar microcavity environment to create photonic modes of sufficient intensity to cause reversible coupling to the exciton dipole moments. Our experiment allows for in-situ tunability of the photon-exciton coupling by variation of the cavity length \cite{flatten_room-temperature_2016}. We analyse the polariton dispersion and probe the state population after off-resonant, optically pumped excitation. These results establish polariton formation in colloidal NPLs, a technological feature that could ne used in broad fields such as laser and semiconductor physics. \\
\noindent
The microcavity consists of two semitransparent silver mirrors ($\approx$ 90\% reflectivity) thermally evaporated onto fused silica substrates. One of the cavity mirrors is on a raised plinth of dimensions $\SI{200}{\micro\meter}$ by $\SI{300}{\micro\meter}$. A concentrated nanoplatelet dispersion is dropcast onto one of the mirrors, forming a dense nanoplatelet film approximately $\SI{700}{\nano\meter}$ thick (see Fig. \ref{fig1}). The mirrors are brought to within $\SI{2}{\micro\meter}$ of each other to form the optical cavity. The cavity length, and hence the cavity photon energy can be scanned in-situ with a piezoelectric actuator attached to one of the cavity mirrors (for details on the experimental setup, see additional information). It is thus possible to access the whole polariton dispersion with different degrees of hybridisation between heavy and light hole excitons.

\section*{\label{results} Results}
The optical absorption spectrum of the nanoplatelets displayed in Fig.~\ref{fig1}b reveals two transitions, which correspond to the lower energy heavy hole (hh) exciton with $E_{\rm{hh}} \approx$ \SI{2.42}{\electronvolt} and the higher-energy light hole (lh) exciton with $E_{\rm{lh}} \approx$ \SI{2.56}{\electronvolt}. These states have binding energies of $R_{\rm{hh}} = (178 \pm 34)$ \SI{}{\milli\electronvolt} and $R_{\rm{hh}} = (259 \pm 3)$ \SI{}{\milli\electronvolt} respectively \cite{naeem_giant_2015}.

To probe the mode structure of the strongly coupled nanoplatelet-cavity system the cavity length $L$ is varied which leads to a shift in the energy of the cavity mode by $\Delta E_{c} = E_{c} \frac{\Delta L}{L}$, where $E_{c} = \frac{qhc}{2L}$ is the energy of the original cavity mode with longitudinal mode number $q$ and $\Delta L$ the variation in cavity length.

\begin{figure*}
\hspace{-20pt}
\includegraphics[width=1.0\textwidth]{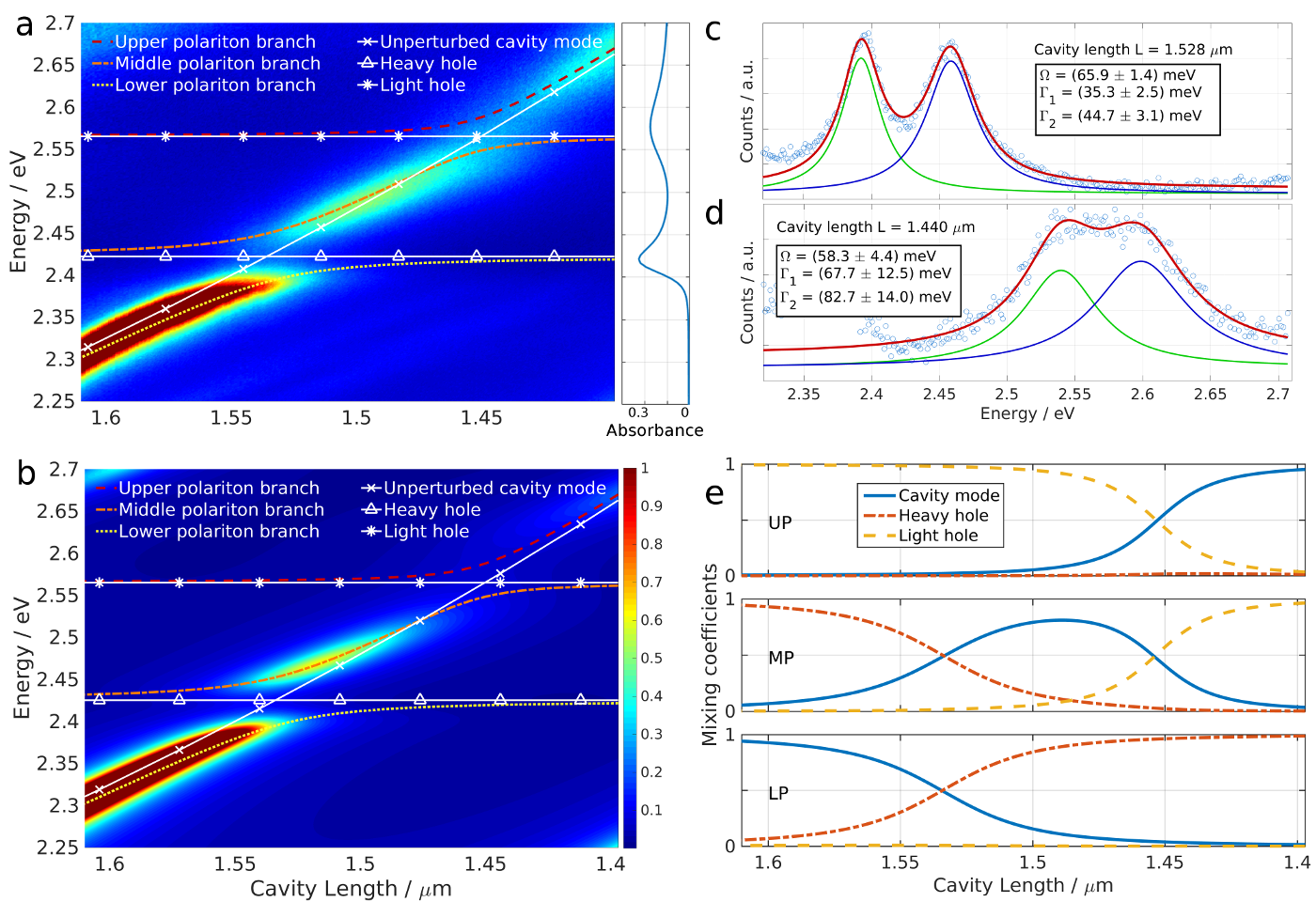}
\caption{Normalised transmission spectra as cavity length is varied, obtained experimentally (a) and by transfer-matrix calculations (b). The bare light and heavy hole exciton energies and the unperturbed cavity mode are overlaid in white, the polariton branches obtained from the Hamiltonian $H$ (Eq. \ref{eq1}) in color. The avoided crossing around the heavy hole transition is resolved fully with a Rabi splitting of $66 \pm \SI{1}{\milli\electronvolt}$, the broad linewidth of the higher energy light hole exciton prevents the resolution of the corresponding splitting, but mode position and intensity clearly indicate the coupling. c) and d) Transmission spectra for cavity lengths $L = \SI{1.528}{\micro\meter}$ and $L = \SI{1.440}{\micro\meter}$ respectively and fits with two Lorentzians revealing the Rabi splitting for the heavy hole  and light hole transition.
e) Square of coefficients $\alpha$, $\beta$ and $\gamma$ showing the hybridisation of photonic mode and excitonic transitions as obtained from diagonalisation of Eq. \ref{eq1}. The three different subplots correspond to the upper (UP), middle (MP) and lower (LP) polariton branches (from top to bottom).
\label{fig2}  }
\end{figure*}

The cavity length dependent optical transmission of the nanoplatelet-cavity system is shown in figure \ref{fig2}a. For the maximum cavity length of $L \approx$ \SI{1.62}{\micro\meter} a transmission peak corresponding to the lower polariton branch (LPB) is visible at \SI{2.3}{\electronvolt}, which moves to higher energy with reducing the cavity length (note the reversed $x$-axis of \ref{fig2}a). As the LPB peak approaches the energy of the hh exciton at \SI{2.42}{\electronvolt}, a second transmission peak, corresponding to the middle polariton branch (MPB), of energy \SI{2.45}{\electronvolt} appears above $E_{\rm{hh}}$. The LPB and MPB energies display an anticrossing about $E_{\rm{hh}}$. As the cavity length is decreased further, a third higher energy transmission peak above $E_{\rm{lh}}$ appears (the upper polariton branch (UPB)) which undergoes an anticrossing with the MPB about $E_{\rm{lh}}$.

This system of one cavity mode simultaneously coupled to two excitonic transitions is described by the Hamiltonian 
\begin{equation}
\begin{split}
H = \  & E_c b^\dagger b
+ E_{\rm{hh}} x_{\rm{hh}}^\dagger x_{\rm{hh}}
+ E_{\rm{lh}} x_{\rm{lh}}^\dagger x_{\rm{lh}} +\\ 
& \  V_{\rm{hh}} (b^\dagger x_{\rm{hh}} + c.c.  )
+ V_{\rm{lh}} (b^\dagger x_{\rm{lh}} + c.c.  )
\end{split}
\label{eq2}
\end{equation}
where $V_{\rm{hh}}$ and $V_{\rm{lh}}$ are the interaction potentials between the cavity mode and heavy and light hole excitons. $b, x_{\rm{hh}}$ and $x_{\rm{lh}}$ are the photon, hh exciton and lh exciton annihilation operators respectively. In the stationary case the system can be reduced to:
\begin{equation}
H \ket{\Psi} 
=
\begin{pmatrix} 
E_{c} & V_{\rm{hh}} & V_{\rm{lh}} \\
V_{\rm{hh}} & E_{\rm{hh}} & 0 \\
V_{\rm{lh}} & 0 & E_{\rm{lh}} 
\end{pmatrix}
\begin{pmatrix} 
\alpha \\
\beta \\
\gamma
\end{pmatrix} = E  \ket{\Psi} 
\label{eq1}
\end{equation}
Here the state $\ket{\Psi}$ is defined by the three coefficients $\alpha$, $\beta$ and $\gamma$, which quantify the contribution of photon, hh exciton and lh exciton respectively.

\begin{figure*}
\hspace{-20pt}
\includegraphics[width=1.0\textwidth]{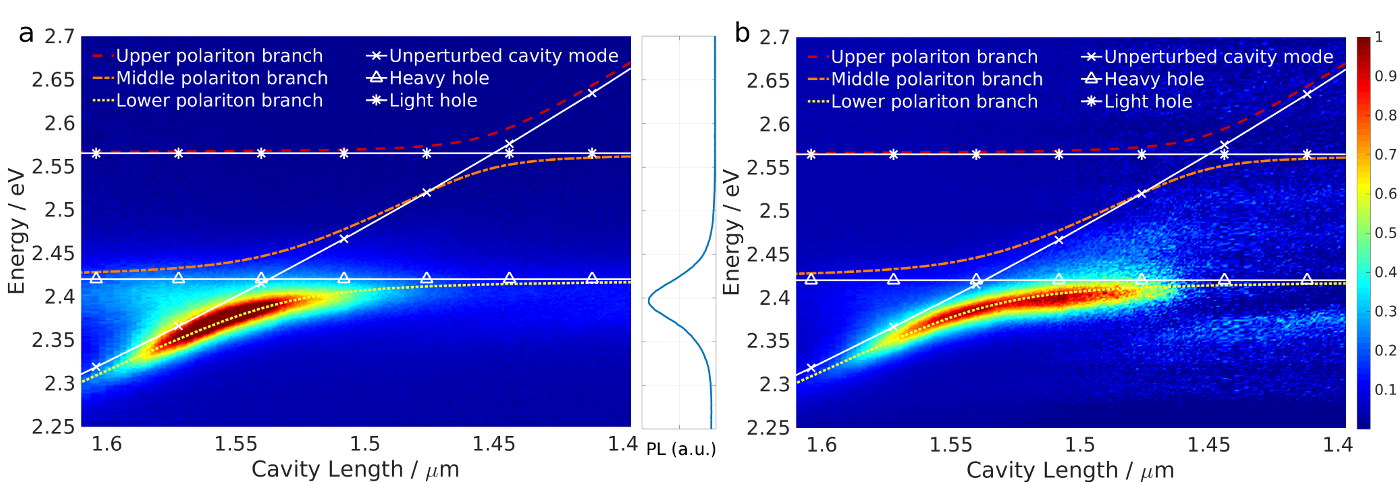}
\caption[flushleft]{a) Photoluminscence intensity of polariton branches as cavity length is varied. Sample is excited off-resonantly with a continuous wave laser with $\lambda = \SI{405}{\nano\meter}$. b) Normalised polariton population obtained from experimental data by scaling each frame with the corresponding inverse of the square of the photonic coefficient $\alpha$. \label{fig3} }
\end{figure*}

Using the known values of $E_{\rm{hh}}$, $E_{\rm{lh}}$ and $E_{c}$, and treating $V_{\rm{hh}}$ and $V_{\rm{lh}}$ as free variables, we fit equation \ref{eq1} to the observed polariton dispersion. At the crossing between unperturbed cavity mode and exciton energy the polaritons are half-light, half-matter quasi-particles. The splitting at this point is the Rabi splitting which corresponds to twice the interaction potential $\hbar \Omega_{hh(lh)} = 2V_{hh(lh)} $. In our system the splitting is well resolved for the heavy hole exciton, where we obtain $\hbar \Omega_{\rm{hh}} = 66 \pm \SI{1}{\milli\electronvolt}$ with linewidths of $\Gamma_{lp} = 35 \pm \SI{3}{\milli\electronvolt}$ and $\Gamma_{mp} = 45 \pm \SI{3}{\milli\electronvolt}$ for lower and middle polariton branch respectively. The light hole transition shows a similar avoided crossing with $\hbar \Omega_{\rm{lh}} = 58 \pm \SI{4}{\milli\electronvolt}$ but the corresponding linewidths of $\Gamma_{mp} = 68 \pm \SI{13}{\milli\electronvolt}$ and $\Gamma_{up} = 83 \pm $\SI{14}{\milli\electronvolt} for middle and upper polariton branch respectively result in a splitting that is not fully resolved. The linewidths and splitting values are obtained by fitting two superposed Lorentzian lineshapes to the transmission spectra at the maximal photon-exciton mixing point (see Fig.~\ref{fig2}c-d), which correspond to vertical cuts through the data presented in Fig.~\ref{fig2}a at cavity lengths $L \approx$ \SI{1.53}{\micro\meter} and $L \approx$ \SI{1.44}{\micro\meter}.

Equation \ref{eq1} allows us to determine the polariton mixing coefficients as a function of cavity length, as shown in Fig.~\ref{fig2}e. For a longitudinal cavity mode with mode number $q$ (in Fig \ref{fig2}a-e we have $q = 6$) the LPB is largely photon like for cavity lengths $L_{cav} >$ \SI{1.53}{\micro\meter}, becoming more hh exciton-like as the cavity mode energy crosses the exciton energy. The UPB similarly is lh exciton-like when the cavity length $L_{cav} >$ \SI{1.44}{\micro\meter}, becoming more photon-like for smaller cavity lengths. Meanwhile, the MPB has hh and lh exciton-like character when close in energy to the hh and lh exciton energies respectively. For cavity lengths \SI{1.53}{\micro\meter} $ > L_{cav} > $ \SI{1.44}{\micro\meter} it has mostly photonic character, however, it also is composed of approximately 10\% of both exciton states, hence it is called a hybridised polariton state. We note that in a transmission experiment, light couples into and out of the cavity via the polariton photon component resulting in greater visibility of photon-like polariton states and observation of a mode structure as shown in Fig.~\ref{fig2}a.

If the number of randomly aligned platelets $N$ coupled to the mode is known, it is possible to deduce the transition dipole moment $\mu$ of a single platelet from the equation \cite{fox_quantum_2006}:
\begin{equation}\label{rabi_splitting}
\hbar \Omega = \frac{1}{3} \mu \sqrt{N} \left( \frac{2\hbar \omega}{\varepsilon_0 n_{\rm{eff}}^2 V_{\rm{eff}}}\right)^{\frac{1}{2}}
\end{equation}
Here $\hbar \omega$ is the exciton energy, $n_{\rm{eff}}$ is the effective refractive index within the cavity and $V_{\rm{eff}}$ is the electric field mode volume. We can then infer the dipole moment of the excitonic transition of a single platelet from the respective Rabi splitting. We find (see. Suppl. Inf.):
\begin{equation}
\begin{split}
  &  N = (3.9 \pm 1.5) \times 10^{5} \\
  &  \mu_{\rm{hh}} = (1.92 \pm 0.37)\times10^{-27} \, \SI{}{\coulomb\meter} = (575 \pm 110) \, \text{D} \\
  &  \mu_{\rm{lh}} = (1.70 \pm 0.35)\times10^{-27} \,  \SI{}{\coulomb\meter} = (509 \pm 104) \, \text{D}
    \end{split}
\end{equation}

This compares to transition dipole moments of $100$~D found in epitaxially grown GaAs quantum dots \cite{guest_measurement_2002} and 
$21 \; \text{to} \; 210$ ~D in InGaN quantum dots with a diameter of \SI{5.2}{\nano\meter} \cite{ostapenko_large_2010}. The lateral dimensions of the platelets studied here are $L_{\rm{x}} = 32.5\pm$\SI{2.5}{\nano\meter} and $L_{\rm{y}} = 8.2\pm$\SI{0.9}{\nano\meter} as obtained by TEM microscopy. The transition dipole moment $\mu$ translates to a lifetime $\tau$ of the heavy hole exciton of $\tau = (1.3 \pm 0.5) $ \SI{}{\pico\second}. Similarly the oscillator strength $f$ of the transition can be found as $f =  (280 \pm 107)$ (see Suppl. Materials Eq. (9) and (10)). These results confirm the giant oscillator strength associated with the large exciton coherence area found in CdSe nanoplatelets \cite{naeem_giant_2015,achtstein_$p$-state_2016}.

Fig.~\ref{fig2}b shows the modelled dispersion curves obtained by Transfer-Matrix-Modelling with the experimentally obtained absorption for the nanoplatelet film. The modelled thickness of the film is $\SI{700}{\nano\meter}$ (see Suppl. Mat. for AFM data) with a peak absorbance of the heavy hole transition of $A = 0.35$, as inferred directly from optical absorption measurement. The absorption curve was converted to complex refractive index values with a classical Lorentz oscillator model\cite{fox_optical_2010} for both transitions, neglecting the continuum of states above $E =$ \SI{2.8}{\electronvolt}. The modelled transmission through the device is in excellent agreement with the experimental data presented in Fig.~\ref{fig2}a.

Fig.~\ref{fig3}a shows the cavity length dependent photoluminescence collected at normal incidence, following non-resonant high energy excitation with a continuous wave laser with $\lambda = \SI{405}{\nano\meter}$. We observe bright emission from the lower polariton branch at cavity lengths between $\SI{1.6}{\micro\meter}$ and $\SI{1.5}{\micro\meter}$, corresponding to a LPB energy of $\SI{2.3}{\milli\electronvolt}$ to $\SI{2.4}{\milli\electronvolt}$. Some weak residual PL is also visible at the energy of the nanoplatelet PL (shown to right of Fig.~\ref{fig3}a). This signal results from the recombination of excitons that are not coupled to the cavity mode, some of which leaks directly though the mirrors. Emission from the MPB and UPB is not visible, as has previously been shown to be a result of fast relaxation pathways between polariton and lower lying exciton states \cite{virgili_polariton_relaxation}. Scaling the photoluminescence intensity with the square of the inverse of the photonic coefficient of the polariton branch allows to see changes in the polariton population along the dispersion, as shown in Fig.~\ref{fig3}b. We note that before scaling the PL intensity to obtain the polariton population, we subtract the uncoupled exciton emission (i.e. the PL spectrum at shortest cavity length). The maximum polariton population occurs at the same energy as the peak exciton emission, suggesting that polariton states are populated primarily through `optical pumping' from uncoupled exciton decay\cite{lidzey_experimental_2002,litinskaya_fast_2004,litinskaya_excitonpolaritons_2006,michetti_exciton-phonon_2009,michetti_simulation_2008}. Here, the radiative decay of excitons directly populates the photonic component of polariton states, however there must be an energetic overlap between the exciton emission and polariton. The small Stokes shift of \SI{12}{\milli\electronvolt} \cite{naeem_giant_2015} therefore limits this population mechanism to polariton states close to the exciton energy.

\section*{\label{conclusion} Conclusion and Outlook}
We have demonstrated the strong coupling between photonic cavity modes and excitons in quasi-two-dimensional colloidal nanoplatelets. The coherent exchange of energy between those two constituents results in the formation of hybridised exciton-polaritons at room temperature with a vacuum Rabi splittings of $66 \pm \SI{1}{\milli\electronvolt}$ and $58 \pm \SI{1}{\milli\electronvolt}$ associated with the heavy hole and light hole exciton transitions respectively. We find that the polariton states are emissive due to the high fluorescence quantum yield of the nanoplatelets, and that polariton states appear to mostly be populated through an optical pumping from uncoupled exciton states. Nanoplatelets represent a promising candidate for polariton based devices due to their large exciton binding energies allowing for room temperature operation. Compared to other colloidal nanoparticles, they are more efficient light absorbers and could become an integral part of future photonic devices.
\section*{Acknowledgements}
We thank Radka Chakalova at the Begbroke Science Park for helping with the thermal evaporation and dicing of the mirrors. L.F. acknowledges funding from the Leverhulme Trust. D.M.C. acknowledges funding from the Oxford Martin School and EPSRC grant EP/K032518/1. The authors declare no competing financial interest.

\putbib

\end{bibunit}

\onecolumngrid
\newpage

\clearpage
\pagebreak
\widetext
\appendix
\begin{center}
\textbf{\large Supplemental Materials: Strong exciton-photon coupling with colloidal nanoplatelets in an open microcavity}
\end{center}
\setcounter{equation}{0}
\setcounter{figure}{0}
\setcounter{table}{0}
\setcounter{page}{1}
\makeatletter
\renewcommand{\theequation}{S\arabic{equation}}
\renewcommand{\thetable}{S\arabic{table}}
\renewcommand{\thefigure}{S\arabic{figure}}
\renewcommand{\bibnumfmt}[1]{[S#1]}
\renewcommand{\citenumfont}[1]{S#1}
\begin{bibunit}
\section{\label{materials} Materials}
Synthesis of CdSe nanoplatelets emitting around \SI{515}{\nano\meter}: In a three-neck round bottom flask 170 mg of Cadmium Myristate, 24 mg of Selenium powder (99,999\%) and 15 ml Octadecene (ODE) were introduced and degassed under vacuum for 30 min at 120 $^{\circ}$C. Then the temperature was raised gradually up to 240 $^{\circ}$C under Argon flow. When the solution turns orange (around 210 $^{\circ}$C) 90 mg of Cadmium Acetate were swiftly added to the reaction. After 10 min of the reaction time the heating mantle was removed and 2.5 ml of Oleic Acid were injected. When the solution had cooled down to 80 $^{\circ}$C 15 ml of Hexane were added. The CdSe nanoplatelets were then separated by selective precipitation.
\\
\newline
Figure 1a shows a TEM micrograph of dispersed platelets. While platelets towards the edges of the agglomeration tend to lay flat of the substrate, other platelets are stacked and lay sideways. The platelets have an average length of $L_x = 32.5 \pm \SI{2.5}{\nano\meter}$, a width of $L_y = 8.2 \pm \SI{0.9}{\nano\meter}$ and a thickness of $L_z =   \SI{1.36}{\nano\meter}$, obtained by taking the inter-plane distance along the 100 direction (13.6 \SI{}{\angstrom}) and 4.5ML of CdSe. Fig. 1b shows the photoluminescence and absorbance lineshape of the platelets.

\section{Methods}
\label{ssec:setup}
The substrate for both sides of the cavity is fused silica. On one side the material is removed with a dicer to form a free standing plinth (Fig. 1f in main text). The plinth and the plain substrate are then coated with a semi-transparent $\SI{50}{\nano\meter}$ silver layer via thermal evaporation. Now the nanoplatelets are dispersed on the main mirror by dropcasting. In the process of drying residual oleic acid acting as a passivating ligand helps in creating a homogeneous film (see Fig. 1d-e in main text).
To form a tunable microcavity the two mirrors are mounted opposite each other, one on a Thorlabs kinematic mount, the other onto a ring piezo-actuator from Piezomechanik. The mirrors are then brought close to each other ($\approx\SI{10}{\micro\meter}$) and made parallel with the help of fabry-perot fringes visible in a transmission experiment with a light emitting diode (Thorlabs MWWHF1). The angular detuning after this procedure is less than \SI{400}{\micro\radian}. The piezo-actuator allows for a length tunability of $\approx \SI{16}{\micro\meter}$ and is driven by a Keithley 2400. Optical access to the sample is given by a standard $\times 10$ ojective lens and the collected light is focused on an Andor combined spectrograph/CCD with a 300 grooves/mm grating. For the photoluminescence experiment the sample is excited with a continuous wave GaN diode laser with $\lambda = \SI{405}{\nano\meter}$ at power densities around $\rho_{exc} = 1000 \frac{W}{cm^2}$.

\section{Elementary analysis in CdSe NPL}

The concentration of the CdSe NPL has been determined with  inductively coupled plasma optical emission spectrometry (ICP-OES). Two measurements of the same sample has been performed and the average has been used to determine the concentration of the CdSe NPL. The results can be summarized in the following table :
\begin{table}[h]
    \centering
    \begin{tabular}{c c c c }
    Element & Avg & Stddev & $\%$RSD  \\ \hline
    Cd$_1$ & 1.304 ppm & 0.003279 & 0.2514 \\
    Se$_1$ & 0.6582 ppm & 0.002482 & 0.3771 \\
    Cd$_2$ & 1.281 ppm & 0.002070 & 0.1616 \\
    Se$_2$ & 0.6536 ppm & 0.001062 &  0.1625 \\ \hline
    Cd & 1.2925 ppm & 0.002674 & 0.2065 \\
    Se & 0.6560 ppm & 0.001772 & 0.2698 \\
    \end{tabular}
    \caption{Results of inductively coupled plasma
optical emission spectrometry (ICP-OES) analysis of CdSe nanoplatelet solution.}
    \label{ICP_table}
\end{table}
\\
The concentration of the CdSe NPL has been determined using the Selenium atoms because the solution has an excess of cadmium oleate.

\section{Measuring the dipole moment}
\begin{figure}[h!]
\centering
\includegraphics[width=1\textwidth]{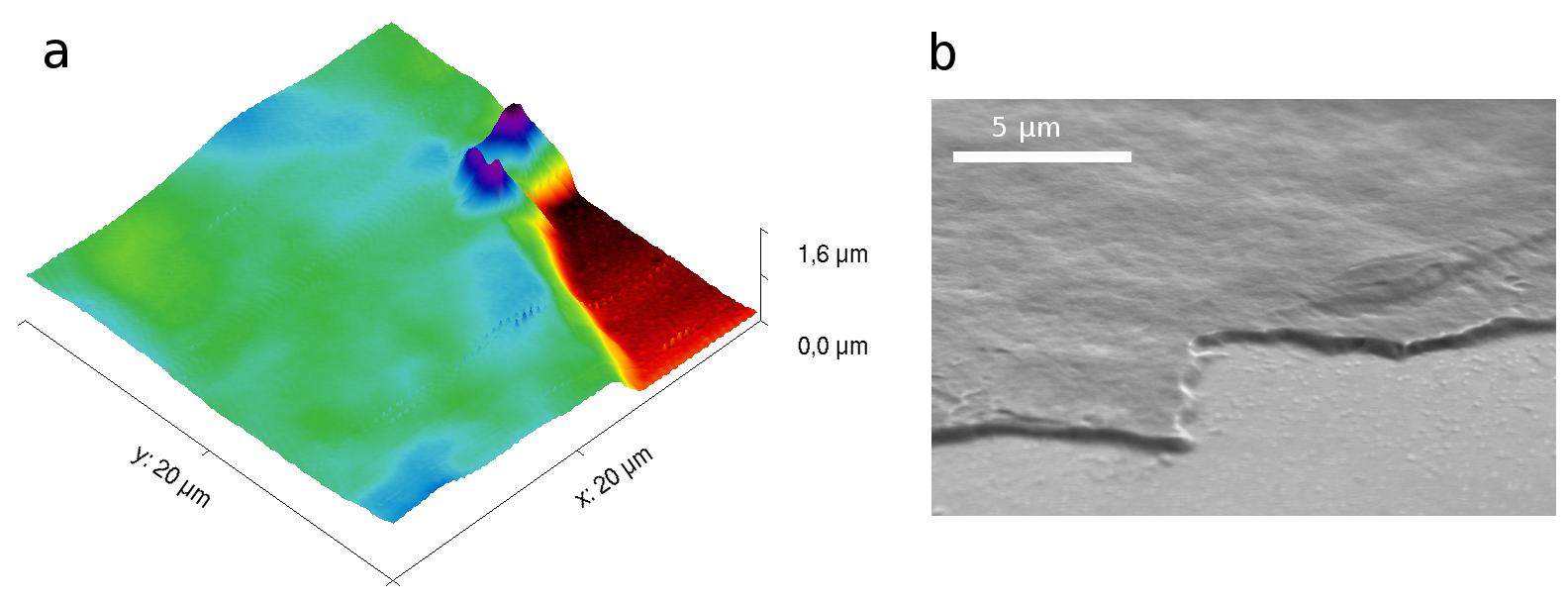}
\caption[flushleft]{a) Atomic Force Microscope image of NPL film close to the edge. b) SEM micrograph of NPL film at an angle of $60^\circ$ to the silver mirror.  \label{fig_suppl_1}}
\end{figure}

In the following we show how to obtain the dipole moment of a single nanoplatelet. The Rabi splitting for an ensemble of $N$ randomly aligned dipoles with transition dipole moment $\mu$ is given as \cite{fox_quantum_2006}:
\begin{equation}\label{rabi_splitting}
\hbar \Omega = \frac{1}{3} \mu \sqrt{N} \left( \frac{2\hbar \omega}{\varepsilon_0 n_{\rm{eff}}^2 V_{\rm{eff}}}\right)^{\frac{1}{2}}
\end{equation}
Here $\hbar \omega$ is the exciton energy, $n_{\rm{eff}}$ is the effective refractive index within the cavity and $V_{\rm{eff}}$ is the electric field mode volume. The factor $\frac{1}{3}$ is obtained by evaluating the integral:
\begin{equation}
\begin{split}
< \vec{\mu} \cdot \hat{E} > & = \frac{1}{4\pi} \int_0^{2\pi} d\phi \int_0^\pi \sin(\theta) d\theta \cos^2(\theta) \mu \\ & = \frac{\mu}{2} \int_0^{\pi} \sin(\theta)\cos^2(\theta) d\theta = \frac{1}{3}\mu
\end{split}
\end{equation} 
We have measured the thickness of the drop casted nanoplatelet film with an AFM  to $d = 693\pm$~\SI{77}{\nano\meter} (see Fig. \ref{fig_suppl_1}a). We deduce the effective refractive index $n_{\rm{eff}} = 1.3$ in accordance with our TMM calculations by taking the weighted average of the refractive index of the film and air within the cavity ($n_{film} = 1.5$, mainly composed of oleic acid). 
To compute $V_{\rm{eff}}$, we take the expression for the effective mode volume for a planar cavity from \cite{ujihara_spontaneous_1991} as
\begin{equation}\label{eff_mode_volume}
V_{\rm{eff}} = \frac{\pi L^2 \lambda}{1-R} \approx 560 \lambda^3 =  \SI{75}{\micro\meter}^3 
\end{equation}
with $\lambda = \frac{2 \pi c}{\omega} =$~\SI{511}{\nano\meter} and the mirror reflectivity $R = 0.95$. 

Now, the only missing parameter is the number of platelets $N$ coupled to the mode. To obtain it, we first measure the average size of nanoplatelets on the basis of a TEM micrograph $L_{\rm{x}} = 32.5\pm$\SI{2.5}{\nano\meter}, $L_{\rm{y}} = 8.2\pm$\SI{0.9}{\nano\meter} and $L_{\rm{z}} = 1.36\pm$\SI{0.05}{\nano\meter}. We then obtain the concentration of selenium atoms in solution by  inductively coupled plasma optical emission spectroscopy (ICP-OES), from which we infer a concentration of $c_{\rm{sol}}^{\rm{NPL}} = 0.890 \pm 0.002~\mu$M. To compare this volume concentration of platelets in solution $c_{\rm{sol}}^{\rm{NPL}}$ to the volume concentration in a dried film $c_{\rm{film}}^{\rm{NPL}}$, we measure the weight difference $\Delta m = $~\SI{6.24}{\milli\gram} of a drop of $V_{\rm{sol}} =(10 \pm 0.18)~$\SI{}{\micro\liter} before and after evaporation $V_{\rm{dry}}$ of the solvent hexane with the known density of $\rho_{\rm{hex}} = 654.8 \frac{\text{mg}}{\text{ml}}$. Note that the uncertainty in the volume pipetted ($\sigma_V = 1.8\%$) dominates the end result. With this we find the volume concentration of platelets in the dried film $c_{\rm{film}}^{\rm{NPL}}$ as:
\begin{equation}
c_{\rm{film}}^{\rm{NPL}} = c_{\rm{sol}}^{\rm{NPL}} \frac{V_{\rm{sol}}}{V_{\rm{dry}}} = c_{\rm{sol}}^{\rm{NPL}} \frac{V_{\rm{sol}}}{V_{\rm{sol}} - \frac{\Delta m}{\rho_{\rm{hex}}}} = 21.26 \, c_{\rm{sol}}^{\rm{NPL}} = 18.9 \pm 6.9 \, \rm{\mu M}
\end{equation}
With the platelet dimensions ($V_{\rm{NPL}} = L_{\rm{x}} L_{\rm{y}} L_{\rm{z}}$) above this concentration translates into a volume fraction $f_V$ of platelets in the film of ($N_A$ is the Avogadro number):
\begin{equation}
f_V = c_{\rm{film}} N_A V_{\rm{NPL}} = ( 0.5 \pm 0.2 )\%
\end{equation}

For the number of platelets coupled to the cavity mode this equates to :
\begin{equation}
N = \frac{V_{eff}}{L} d c_{\rm{film}}^{\rm{NPL}} N_A = \frac{\pi L \lambda}{1-R} d c_{\rm{film}}^{\rm{NPL}} N_A = (3.9 \pm 1.5) \times 10^{5}
\end{equation}
Taking this value and using $\hbar \Omega = (65.9 \pm 1.4)\, \SI{}{\milli\electronvolt}$ we solve Eq. \ref{rabi_splitting} for $\mu$ and obtain:
\begin{equation}
\mu = 3  \left( \frac{\varepsilon_0 n_{\rm{eff}}^2 V_{\rm{eff}}}{2\hbar \omega}\right)^{\frac{1}{2}} \frac{\hbar \Omega}{\sqrt{N}} = (1.92 \pm 0.37)\times10^{-27} \, \SI{}{\coulomb\meter} = (575 \pm 110) \, \text{D}
\end{equation}
Note that the value for $\mu$ is independent of the exact choice for the mode volume (and thus the mirror reflectivity) since:
\small{
\begin{equation}
\begin{split}
\mu & = 3  \left( \frac{\varepsilon_0 n_{\rm{eff}}^2 V_{\rm{eff}}}{2\hbar \omega}\right)^{\frac{1}{2}} \frac{\hbar \Omega}{\sqrt{N}} = 3  \left( \frac{\varepsilon_0 n_{\rm{eff}}^2 V_{\rm{eff}}}{2\hbar \omega}\right)^{\frac{1}{2}} \frac{\hbar \Omega \sqrt{L} }{\sqrt{V_{\rm{eff}} d c_{\rm{film}}^{\rm{NPL}} N_A }} \\ & = 3  \left( \frac{\varepsilon_0 n_{\rm{eff}}^2 }{2\hbar \omega  c_{\rm{film}}^{\rm{NPL}} N_A}\right)^{\frac{1}{2}} \left( \frac{L}{d}\right)^{\frac{1}{2}} \hbar \Omega
\end{split}
\end{equation}}
Assuming a two level system with energy difference $\Delta E = \hbar \omega$ and transition dipole moment $\mu$, the radiative lifetime $\tau$ of the heavy-hole exciton can be expressed as \cite{fox_quantum_2006}: 
\begin{equation}
\tau = \frac{3 \pi \epsilon_0 \hbar c^3}{\omega^3 \mu^2}
\end{equation}
With $\mu$ as found above this expression results in $\tau = (1.3 \pm 0.5) $ \SI{}{\pico\second}. Similarly the oscillator strength $f$ of the transition can be expressed as:
\begin{equation}
f = \frac{2 m \omega}{3 \hbar} \mu^2 = (280 \pm 107)
\end{equation}
where $m$ is the reduced mass of the exciton. These results confirm the giant oscillator strength associated with the large exciton coherence area reported previously \cite{naeem_giant_2015}.

\putbib

\end{bibunit}


\begin{thebibliography}{10}

\bibitem{weisbuch_observation_1992}
C.~Weisbuch, M.~Nishioka, A.~Ishikawa, and Y.~Arakawa, ``Observation of the
  coupled exciton-photon mode splitting in a semiconductor quantum
  microcavity,'' {\em Physical Review Letters}, vol.~69, pp.~3314--3317, Dec.
  1992.

\bibitem{lidzey_strong_1998}
D.~G. Lidzey, D.~D.~C. Bradley, M.~S. Skolnick, T.~Virgili, S.~Walker, and
  D.~M. Whittaker, ``Strong exciton–photon coupling in an organic
  semiconductor microcavity,'' {\em Nature}, vol.~395, pp.~53--55, Sept. 1998.

\bibitem{agranovich_cavity_2003}
V.~M. Agranovich, M.~Litinskaia, and D.~G. Lidzey, ``Cavity polaritons in
  microcavities containing disordered organic semiconductors,'' {\em Physical
  Review B}, vol.~67, p.~085311, Feb. 2003.

\bibitem{dufferwiel_exciton-polaritons_2015}
S.~Dufferwiel, S.~Schwarz, F.~Withers, A.~a.~P. Trichet, F.~Li, M.~Sich, O.~Del
  Pozo-Zamudio, C.~Clark, A.~Nalitov, D.~D. Solnyshkov, G.~Malpuech, K.~S.
  Novoselov, J.~M. Smith, M.~S. Skolnick, D.~N. Krizhanovskii, and A.~I.
  Tartakovskii, ``Exciton-polaritons in van der {Waals} heterostructures
  embedded in tunable microcavities,'' {\em Nature Communications}, vol.~6,
  p.~8579, Oct. 2015.

\bibitem{liu_strong_2015}
X.~Liu, T.~Galfsky, Z.~Sun, F.~Xia, E.-c. Lin, Y.-H. Lee, S.~K\'{e}na-Cohen,
  and V.~M. Menon, ``Strong light–matter coupling in two-dimensional atomic
  crystals,'' {\em Nature Photonics}, vol.~9, pp.~30--34, Jan. 2015.

\bibitem{giebink_strong_2011}
N.~C. Giebink, G.~P. Wiederrecht, and M.~R. Wasielewski, ``Strong
  exciton-photon coupling with colloidal quantum dots in a high-{Q} bilayer
  microcavity,'' {\em Applied Physics Letters}, vol.~98, p.~081103, Feb. 2011.

\bibitem{christopoulos_room-temperature_2007}
S.~Christopoulos, G.~B.~H. von H\"{o}gersthal, A.~J.~D. Grundy, P.~G.
  Lagoudakis, A.~V. Kavokin, J.~J. Baumberg, G.~Christmann, R.~Butt\'{e},
  E.~Feltin, J.-F. Carlin, and N.~Grandjean, ``Room-{Temperature} {Polariton}
  {Lasing} in {Semiconductor} {Microcavities},'' {\em Physical Review Letters},
  vol.~98, p.~126405, Mar. 2007.

\bibitem{kena-cohen_room-temperature_2010}
S.~K\'{e}na-Cohen and S.~R. Forrest, ``Room-temperature polariton lasing in an
  organic single-crystal microcavity,'' {\em Nature Photonics}, vol.~4,
  pp.~371--375, June 2010.

\bibitem{sich_observation_2012}
M.~Sich, D.~N. Krizhanovskii, M.~S. Skolnick, A.~V. Gorbach, R.~Hartley, D.~V.
  Skryabin, E.~A. Cerda-M\'{e}ndez, K.~Biermann, R.~Hey, and P.~V. Santos,
  ``Observation of bright polariton solitons in a semiconductor microcavity,''
  {\em Nature Photonics}, vol.~6, pp.~50--55, Jan. 2012.

\bibitem{amo_superfluidity_2009}
A.~Amo, J.~Lefr\`{e}re, S.~Pigeon, C.~Adrados, C.~Ciuti, I.~Carusotto,
  R.~Houdr\'{e}, E.~Giacobino, and A.~Bramati, ``Superfluidity of polaritons in
  semiconductor microcavities,'' {\em Nature Physics}, vol.~5, pp.~805--810,
  Nov. 2009.

\bibitem{byrnes_exciton-polariton_2014}
T.~Byrnes, N.~Y. Kim, and Y.~Yamamoto, ``Exciton-polariton condensates,'' {\em
  Nature Physics}, vol.~10, pp.~803--813, Nov. 2014.

\bibitem{ekimov_quantum_1985}
A.~I. Ekimov, A.~L. Efros, and A.~A. Onushchenko, ``Quantum size effect in
  semiconductor microcrystals,'' {\em Solid State Communications}, vol.~56,
  pp.~921--924, Dec. 1985.

\bibitem{yin_colloidal_2005}
Y.~Yin and A.~P. Alivisatos, ``Colloidal nanocrystal synthesis and the
  organic–inorganic interface,'' {\em Nature}, vol.~437, pp.~664--670, Sept.
  2005.

\bibitem{guyot-sionnest_colloidal_2008}
P.~Guyot-Sionnest, ``Colloidal quantum dots,'' {\em Comptes Rendus Physique},
  vol.~9, pp.~777--787, Oct. 2008.

\bibitem{carbone_synthesis_2007}
L.~Carbone, C.~Nobile, M.~De~Giorgi, F.~D. Sala, G.~Morello, P.~Pompa,
  M.~Hytch, E.~Snoeck, A.~Fiore, I.~R. Franchini, M.~Nadasan, A.~F. Silvestre,
  L.~Chiodo, S.~Kudera, R.~Cingolani, R.~Krahne, and L.~Manna, ``Synthesis and
  {Micrometer}-{Scale} {Assembly} of {Colloidal} {CdSe}/{CdS} {Nanorods}
  {Prepared} by a {Seeded} {Growth} {Approach},'' {\em Nano Letters}, vol.~7,
  pp.~2942--2950, Oct. 2007.

\bibitem{ithurria_quasi_2008}
S.~Ithurria and B.~Dubertret, ``Quasi 2d {Colloidal} {CdSe} {Platelets} with
  {Thicknesses} {Controlled} at the {Atomic} {Level},'' {\em Journal of the
  American Chemical Society}, vol.~130, pp.~16504--16505, Dec. 2008.

\bibitem{ithurria_colloidal_2011}
S.~Ithurria, M.~D. Tessier, B.~Mahler, R.~P. S.~M. Lobo, B.~Dubertret, and
  A.~L. Efros, ``Colloidal nanoplatelets with two-dimensional electronic
  structure,'' {\em Nature Materials}, vol.~10, pp.~936--941, Dec. 2011.

\bibitem{tessier_spectroscopy_2012}
M.~D. Tessier, C.~Javaux, I.~Maksimovic, V.~Loriette, and B.~Dubertret,
  ``Spectroscopy of {Single} {CdSe} {Nanoplatelets},'' {\em ACS Nano}, vol.~6,
  pp.~6751--6758, Aug. 2012.

\bibitem{naeem_giant_2015}
A.~Naeem, F.~Masia, S.~Christodoulou, I.~Moreels, P.~Borri, and W.~Langbein,
  ``Giant exciton oscillator strength and radiatively limited dephasing in
  two-dimensional platelets,'' {\em Physical Review B}, vol.~91, p.~121302,
  Mar. 2015.

\bibitem{flatten_room-temperature_2016}
L.~C. Flatten, Z.~He, D.~M. Coles, A.~A.~P. Trichet, A.~W. Powell, R.~A.
  Taylor, J.~H. Warner, and J.~M. Smith, ``Room-temperature exciton-polaritons
  with two-dimensional {WS}$_2$,'' {\em arXiv:1605.04743 [cond-mat,
  physics:physics]}, May 2016.
\newblock arXiv: 1605.04743.

\bibitem{fox_quantum_2006}
M.~Fox, {\em Quantum {Optics}: {An} {Introduction}}.
\newblock OUP Oxford, Apr. 2006.

\bibitem{guest_measurement_2002}
J.~R. Guest, T.~H. Stievater, X.~Li, J.~Cheng, D.~G. Steel, D.~Gammon, D.~S.
  Katzer, D.~Park, C.~Ell, A.~Thr\"{a}nhardt, G.~Khitrova, and H.~M. Gibbs,
  ``Measurement of optical absorption by a single quantum dot exciton,'' {\em
  Physical Review B}, vol.~65, p.~241310, June 2002.

\bibitem{ostapenko_large_2010}
I.~A. Ostapenko, G.~H\"{o}nig, C.~Kindel, S.~Rodt, A.~Strittmatter,
  A.~Hoffmann, and D.~Bimberg, ``Large internal dipole moment in {InGaN}/{GaN}
  quantum dots,'' {\em Applied Physics Letters}, vol.~97, p.~063103, Aug. 2010.

\bibitem{achtstein_$p$-state_2016}
A.~W. Achtstein, R.~Scott, S.~Kickh\"{o}fel, S.~T. Jagsch, S.~Christodoulou,
  G.~H. Bertrand, A.~V. Prudnikau, A.~Antanovich, M.~Artemyev, I.~Moreels,
  A.~Schliwa, and U.~Woggon, ``\$p\$-{State} {Luminescence} in {CdSe}
  {Nanoplatelets}: {Role} of {Lateral} {Confinement} and a {Longitudinal}
  {Optical} {Phonon} {Bottleneck},'' {\em Physical Review Letters}, vol.~116,
  p.~116802, Mar. 2016.

\bibitem{fox_optical_2010}
M.~Fox, {\em Optical {Properties} of {Solids}, Ch. 2}.
\newblock Oxford University Press, Mar. 2010.

\bibitem{virgili_polariton_relaxation}
T.~Virgili, D.~Coles, A.~M. Adawi, C.~Clark, P.~Michetti, S.~K. Rajendran,
  D.~Brida, D.~Polli, G.~Cerullo, and D.~G. Lidzey, ``Ultrafast polariton
  relaxation dynamics in an organic semiconductor microcavity,'' {\em Phys.
  Rev. B}, vol.~83, p.~245309, Jun 2011.

\bibitem{lidzey_experimental_2002}
D.~G. Lidzey, A.~M. Fox, M.~D. Rahn, M.~S. Skolnick, V.~M. Agranovich, and
  S.~Walker, ``Experimental study of light emission from strongly coupled
  organic semiconductor microcavities following nonresonant laser excitation,''
  {\em Physical Review B}, vol.~65, p.~195312, May 2002.

\bibitem{litinskaya_fast_2004}
M.~Litinskaya, P.~Reineker, and V.~M. Agranovich, ``Fast polariton relaxation
  in strongly coupled organic microcavities,'' {\em Journal of Luminescence},
  vol.~110, pp.~364--372, Dec. 2004.

\bibitem{litinskaya_excitonpolaritons_2006}
M.~Litinskaya, P.~Reineker, and V.~M. Agranovich, ``Exciton–polaritons in
  organic microcavities,'' {\em Journal of Luminescence}, vol.~119-120,
  pp.~277--282, July 2006.

\bibitem{michetti_exciton-phonon_2009}
P.~Michetti and G.~C. La~Rocca, ``Exciton-phonon scattering and photoexcitation
  dynamics in \${J}\$-aggregate microcavities,'' {\em Physical Review B},
  vol.~79, p.~035325, Jan. 2009.

\bibitem{michetti_simulation_2008}
P.~Michetti and G.~C. La~Rocca, ``Simulation of {J}-aggregate microcavity
  photoluminescence,'' {\em Physical Review B}, vol.~77, p.~195301, May 2008.

\end{thebibliography}


\begin{thebibliography}{1}

\bibitem{fox_quantum_2006}
M.~Fox, {\em Quantum {Optics}: {An} {Introduction}}.
\newblock OUP Oxford, Apr. 2006.

\bibitem{ujihara_spontaneous_1991}
K.~Ujihara, ``Spontaneous {Emission} and the {Concept} of {Effective} {Area} in
  a {Very} {Short} {Optical} {Cavity} with {Plane}-{Parallel} {Dielectric}
  {Mirrors},'' {\em Japanese Journal of Applied Physics}, vol.~30,
  pp.~L901--L903, May 1991.

\bibitem{naeem_giant_2015}
A.~Naeem, F.~Masia, S.~Christodoulou, I.~Moreels, P.~Borri, and W.~Langbein,
  ``Giant exciton oscillator strength and radiatively limited dephasing in
  two-dimensional platelets,'' {\em Physical Review B}, vol.~91, p.~121302,
  Mar. 2015.

\end{thebibliography}
\end{document}